\documentclass[aip,jap,unsortedaddress,superscriptaddress,reprint]{revtex4-1}
\usepackage{hyperref}
\usepackage{graphicx}
\usepackage{amsmath}
\usepackage{amssymb}
\usepackage{times,mathptmx}
\usepackage{dcolumn}
\usepackage{ifthen}
\usepackage{color}
\usepackage{proof}

\newif\ifcom
\newif\ifdel
\comtrue               %
\delfalse

\begin{document}

\title{Investigation of the tunnel magnetoresistance in junctions with a strontium stannate barrier}
\author{Matthias Althammer}
\email{Matthias.Althammer@wmi.badw.de}
\affiliation{MINT Center, University of Alabama, Tuscaloosa AL 35487 USA}
\affiliation{Walther-Mei{\ss}ner-Institut, Bayerische Akademie der Wissenschaften, 85748 Garching, Germany}
\author{Amit Vikam Singh}
\affiliation{MINT Center, University of Alabama, Tuscaloosa AL 35487 USA}
\affiliation{Department of Chemistry and Chemical Engineering, University of Alabama, Tuscaloosa AL 35487 USA}
\author{Sahar Keshavarz}
\affiliation{MINT Center, University of Alabama, Tuscaloosa AL 35487 USA}
\affiliation{Department of Physics and Astronomy, University of Alabama, Tuscaloosa AL 35487 USA}
\author{Mehmet Kenan Yurtisigi}
\affiliation{MINT Center, University of Alabama, Tuscaloosa AL 35487 USA}
\affiliation{Department of Physics and Astronomy, University of Alabama, Tuscaloosa AL 35487 USA}
\author{Rohan Mishra}
\affiliation{Department of Mechanical Engineering and Materials Science, Washington University in St. Louis, St. Louis, MO 63130 USA}
\affiliation{ Materials Sciences and Technology Division, Oak Ridge National Laboratory, Oak Ridge,TN 37831 USA}
\author{Albina Borisevich}
\affiliation{ Materials Sciences and Technology Division, Oak Ridge National Laboratory, Oak Ridge,TN 37831 USA}
\author{Patrick LeClair}
\affiliation{MINT Center, University of Alabama, Tuscaloosa AL 35487 USA}
\affiliation{Department of Physics and Astronomy, University of Alabama, Tuscaloosa AL 35487 USA}
\author{Arunava Gupta}
\affiliation{MINT Center, University of Alabama, Tuscaloosa AL 35487 USA}
\affiliation{Department of Chemistry and Chemical Engineering, University of Alabama, Tuscaloosa AL 35487 USA}
\date{\today}
\begin{abstract}
We experimentally investigate the structural, magnetic and electrical transport properties of La$_{0.67}$Sr$_{0.33}$MnO$_3$ based magnetic tunnel junctions with a SrSnO$_3$ barrier. Our results show that despite the large number of defects in the strontium stannate barrier, due to the large lattice mismatch, the observed tunnel magnetoresistance is comparable to tunnel junctions with a better lattice matched STiO$_3$ barrier, reaching values of up to $350\;\%$ at $T=5\;\mathrm{K}$. Further analysis of the current-voltage characteristics of the junction and the bias voltage dependence of the observed tunnel magnetoresistance show a decrease of the TMR with increasing bias voltage. In addition, the observed TMR vanishes for $T>200\;\mathrm{K}$. Our results suggest that by employing a better lattice matched ferromagnetic electrode and thus reducing the structural defects in the strontium stannate barrier even larger TMR ratios might be possible in the future.
\end{abstract}
\maketitle
\section{Introduction}
Electric tunnel junctions consist of two electrically conductive electrodes separated by an insulating tunnel barrier. In these tunnel junctions the barrier plays a crucial role for optimizing and engineering device properties. For example, by employing a barrier with a ferroelectric polarization it is possible to exploit the polarization state as a means to store information in the so called tunnel electroresistance states~\cite{Garcia2014,Kim2013,Singh2015}, or utilizing a ferromagnetic barrier for spin filtering properties~\cite{Moodera_2007}. In magnetic tunnel junctions (MTJs), consisting of two ferromagnetic electrodes, the tunnel magnetoresistance (TMR) can be further enhanced by utilizing wave function filtering of the tunnel barrier, as for example in Fe/MgO/Fe tunnel junctions with TMR ratios reaching up to $800\;\%$ at room temperature~\cite{Zhang_Butler_2004,Parkin2004,Yuasa2004,Ikeda_TMR_2008}. However, for large TMR ratios not only the functional properties are relevant, but also a good lattice match between the electrode and barrier materials are necessary. This problem becomes apparent in MTJs based on Heusler electrodes with MgO barrier, where the lattice mismatch between electrodes and barrier reduces significantly the observed TMR~\cite{ReviewHeusler_2012}.

In the following we investigate ferromagnetic La$_{0.67}$Sr$_{0.33}$MnO$_{3}$(LSMO) based MTJs utilizing a novel insulating barrier material: strontium tin oxide (SrSnO$_3$, SSO). SSO crystalizes in a cubic perovskite structure with a lattice constant $a=0.40254\,\mathrm{nm}$ (Ref.~\onlinecite{SSO-crystal}). Up to now SSO has only been used as a transparent conductive layer, when doped with La, Ba,~\cite{liu_structure_2012,kurre_studies_2011,wang_transparent_2010} as an insulation barrier with a high dielectric constant in single flux quantum circuits~\cite{wakana_examination_2005}, and as a photoelectrochemical converter for the reduction of water~\cite{bellal_visible_2009}. Our experiments show that it is possible to fabricate MTJs with an SSO barrier that exhibt large TMR ratios of up to $350\:\%$ at liquid He temperatures and large resistance-area products of up to $30\;\mathrm{M\Omega \mu m^2}$. We compare these results to reference MTJs with a SrTiO$_{3}$ (STO) barrier and compare the obtained results to already existing publications. We start with a short description of the experimental techniques used, then present the results of our experiments and conclude with a summary.

\section{Experiment}

The samples investigated here have been grown by pulsed laser deposition (PLD) from stoichiometric, ceramic LSMO, SSO and STO targets (99.99\% purity) on (001)-oriented STO substrates. The pulsed laser deposition was carried out in an UHV-chamber with a base pressure of $2\times10^{-7}\;\mathrm{Torr}$ using a KrF excimer laser ($248\;\mathrm{nm}$ wavelength, $2\;\mathrm{Hz}$ repetition rate) at a substrate temperature of $700^\circ\mathrm{C}$. For the LSMO, SSO, and STO deposition a laser fluence of $1.2\;\mathrm{J/cm^2}$, $0.8\;\mathrm{J/cm^2}$, and $1.2\;\mathrm{J/cm^2}$ and an oxygen atmosphere $200\;\mathrm{mTorr}$, $200\;\mathrm{mTorr}$, and $100\;\mathrm{mTorr}$, respectively, has been used. After deposition the samples have been cooled down to room temperature in an oxygen atmosphere of $200\;\mathrm{mTorr}$

To determine the crystal phase and epitaxy of the resulting films, a standard 4 circle x-ray diffraction (XRD) setup (Phillips X’pert Pro) was used with a Cu $K\alpha$ source. Transmission electron microscope (TEM) cross sectional samples were prepared by conventional mechanical polishing and Ar ion milling. The scanning TEM (STEM) imaging were carried out in an aberration-corrected Nion UltraSTEM 200 microscope operating at $200\;\mathrm{kV}$.

For the fabrication of the MTJs from the blanket films we utilized a three step photolithography process as detailed in Ref.~\onlinecite{lu_large_1996}. In the first step, we defined the bottom contact mesa by an initial Ar ion milling etch followed by a chemical wet etch in diluted HCl (1:10) to avoid Ar-ion induced conductivity of the STO substrate.~\cite{gross_situ_2011} After the etching process, a SiO$_2$ insulation layer was sputter deposited and the photoresist removed via lift-off. In the second step the tunnel junctions were formed by Ar ion milling and SiO$_2$ deposition. In the final step, top contacts were fabricated by Ru sputter deposition and lift-off.

For the electrical characterization the samples were mounted in the variable temperature insert of a Dynacool PPMS system at temperatures $5\;\mathrm{K} \leq T \leq 300\;\mathrm{K}$ and in magnetic fields $H$ applied in the film plane of up to $9\;\mathrm{T}$. A constant DC-bias voltage was applied to the MTJs, while recording the 4-pt DC-voltage drop $V_\mathrm{4pt}$ using an Agilent HP 3478 voltmeter and the DC current flow $I$ using a Keithley 428 current amplifier and an Agilent HP 3458a voltmeter across the junction (See insert in Fig.~\ref{figure:RT_LSMO_SSO}).

\section{Results and Discussion}

\begin{figure}[t,b]
  \includegraphics[width=85mm]{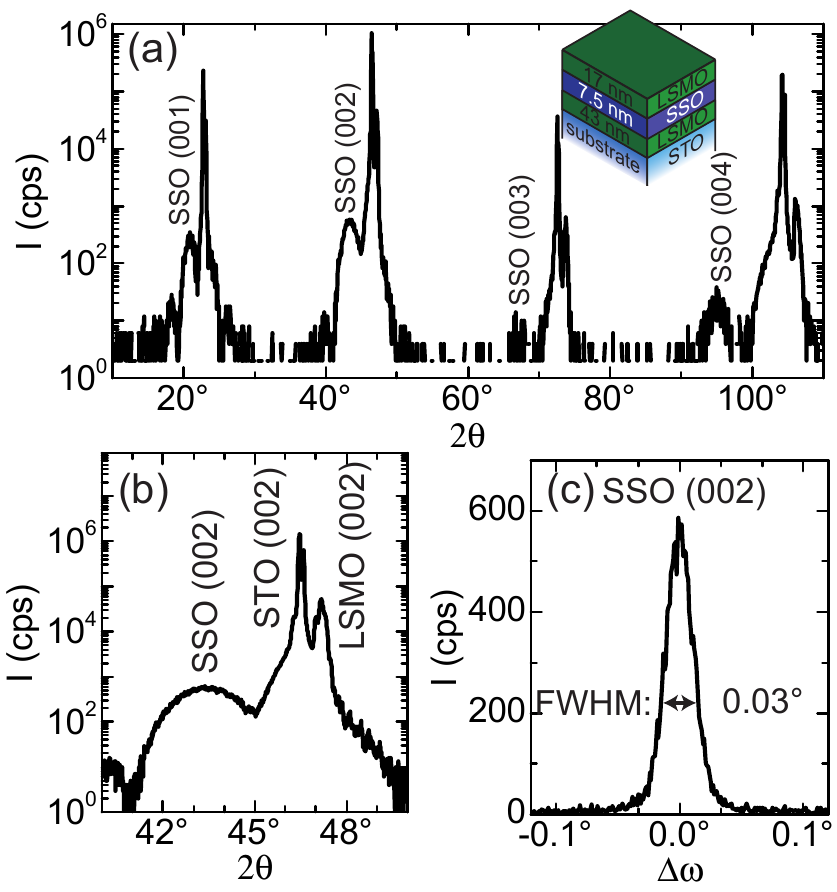}\\
  \caption[XRD results for LSMO based Tunneljunctions]{(Color online) (a) Full range $2\theta-\omega$-scan for a LSMO(43)/SSO(7.5)/LSMO(17) trilayer on a STO (001) substrate. Only $(00l)$ reflections from STO, LSMO, and SSO are visible. (b) Enlarged scale of the $2\theta-\omega$-scan of (a) around the STO (002) reflection. (c) Rocking curve of the SSO(002) reflection exhibits a FWHM of $0.03^\circ$ indicating a small mosaic spread.
  \label{figure:XRD_LSMO}}
\end{figure}

We first discuss the XRD data exemplarily obtained for a LSMO(43)/SSO(7.5)/LSMO(17) trilayer; here as throughout the text values in parenthesis are the corresponding film thicknesses in nm. The full range $2\theta-\omega$-scan in Fig.~\ref{figure:XRD_LSMO}(a) exhibits no secondary phases. Only peaks that can be related to $(00l)$-reflections of the LSMO, SSO layers and the STO substrate are visible and even Laue oscillations around the SSO $(001)$-reflection are present. This indicates a good epitaxial (001)-oriented growth of SSO on LSMO. Moreover, in the high resolution $2\theta-\omega$-scan around the STO $(002)$-reflection the characteristic $K\alpha_1$, $K\alpha_2$ peak splitting of the LSMO peak indicates a high crystalline quality of the LSMO layers. A rocking curve of the LSMO $(002)$-reflection (not shown here) yields a full width at half maximum (FWHM) of the reflection peak of $0.02^\circ$, which is close to the instrumental resolution of our setup. From the peak position of the SSO $(002)$-reflection we extract an out-of-plane lattice constant of $0.413\;\mathrm{nm}$, that is slightly larger then the bulk value of SSO, which we attribute to the compressive strain due to the growth on LSMO on a STO substrate (bulk lattice constant LSMO $a_\mathrm{LSMO}=0.3894\;\mathrm{nm}$, bulk lattice constant STO $a_\mathrm{STO}=0.3902\;\mathrm{nm}$, Ref.~\onlinecite{Vailionis2011}). For the rocking curve of the SSO $(002)$-reflection shown in Fig.~\ref{figure:XRD_LSMO}(c) we obtain a FWHM of $0.03^\circ$, indicating a low mosaic spread of the SSO layer. We also employed $\phi$-scans at the (101) and (202) reflections of substrate and deposited layers (data not shown here), which verified the in-plane epitaxial relation $[100]_\mathrm{STO}\parallel[100]_\mathrm{SSO}\parallel[100]_\mathrm{LSMO}$. Taking all these findings together, our deposition parameters yield highly epitaxial LSMO/SSO/LSMO trilayers with excellent structural quality. Similar results have also been obtained with the very same growth conditions for our reference LSMO/STO/LSMO trilayers.

\begin{figure*}[b,t]
  \includegraphics[width=170mm]{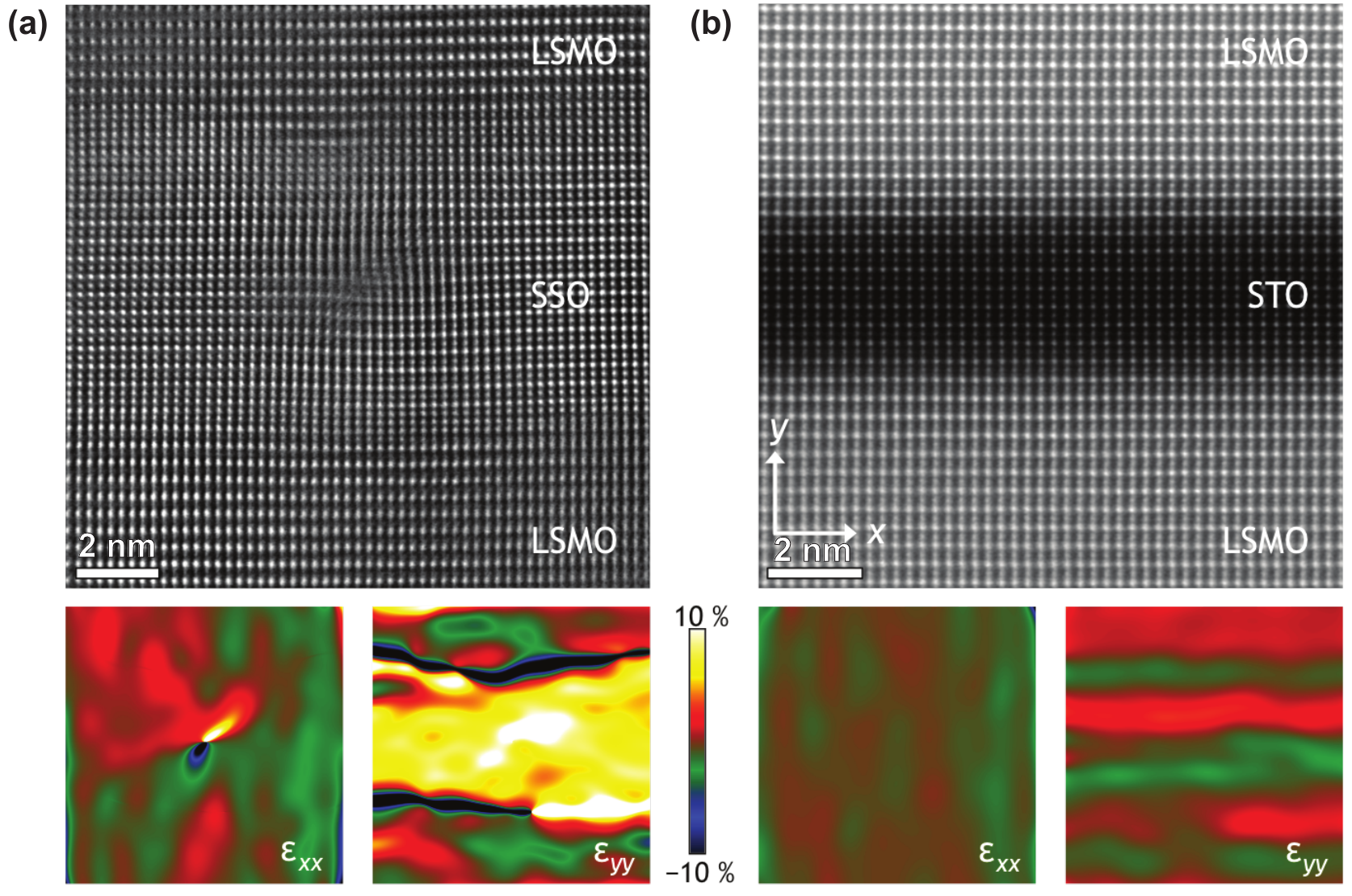}\\
  \caption[STEM images of SSO and STO trilayers]{(Color online)STEM Z-contrast images of $[110]_\mathrm{pc}$-oriented (a) LSMO(43)/SSO(7.5)/LSMO(17) trilayer showing a dislocation in the SSO layer and (b) LSMO(50)/STO(4)/LSMO(30) trilayer. The colormaps below the Z-contrast images show the distribution of strain tensor components $\epsilon_\mathrm{xx}$ and $\epsilon_\mathrm{yy}$ in the respective images, where x and y are the in-plane and out-of-plane directions as shown by the arrows in (b). The scale bar in both the images is $2\;\mathrm{nm}$.
  \label{figure:TEM_LSMO}}
\end{figure*}

We now focus on the comparison of the structural quality of We now focus on the comparison of the structural quality of the two trilayers obtained from atomic resolution STEM. Figure~\ref{figure:TEM_LSMO}(a) and (b) show high-angle annular dark field (HAADF) STEM images obtained for a LSMO(43)/SSO(7.5)/LSMO(17) and a LSMO(50)/STO(4)/LSMO(30) trilayer, respectively. Both the films are oriented along the $[110]_\mathrm{pc}$ direction (pc denotes pseudocubic). As the contrast in a HAADF STEM image is roughly proportional to $Z^2$ (Ref.~\onlinecite{Pennycook_2011}), with Z being the atomic number, the cation columns in both the trilayers can be clearly observed, while the lighter oxygen columns are invisible. Within the LSMO layers, the (La/Sr)O layers appear brighter than the MnO$_2$ layers. Similarly in the STO barrier in Fig.~\ref{figure:TEM_LSMO}(b), SrO layers appear brighter than TiO$_2$ layers. However, In the SSO barrier in Fig.~\ref{figure:TEM_LSMO}(a), Sn (Z = 50) being heavier than Sr (Z = 38), the SnO$_2$ layers appear brighter than the SrO layers. While it is clear that the LSMO/STO/LSMO trilayer in Fig.~\ref{figure:TEM_LSMO}(a) is free of any line defects and is of a high quality, the SSO layer and the LSMO/SSO interfaces are defective with the presence of dislocations. We performed geometric phase analysis (GPA) of the two Z-contrast images to quantify the strain within the trilayers \cite{Htch1998}, which are shown as colormaps below the respective images. For the LSMO/SSO/LSMO trilayer, GPA of $\epsilon_\mathrm{yy}$ component of the stress tensor shows presence of a very large dilation ($\approx 9\% $) in the out-of-plane direction in the SSO barrier, as expected due to the larger lattice constant of SSO. The $\epsilon_\mathrm{xx}$ component on the other hand shows large in-plane dilation and compression around the core of the dislocation, which is formed due to the lattice mismatch. In contrast, GPA of the STO barrier reveals only a small strain along the growth direction.
From the TEM analysis we can conclude that the SSO barrier layer contains a higher density of defects, due to the larger lattice misfit between SSO and LSMO. This seems to be in contrast to our structural analysis via XRD, where we found excellent structural quality. This difference can be rationalized by the integral properties of XRD diffraction, while TEM reflects local properties of a sample. This highlights the importance of microstructural analysis via TEM imaging when analyzing the structural quality of samples.

\begin{figure}[b,t]
  \includegraphics[width=85mm]{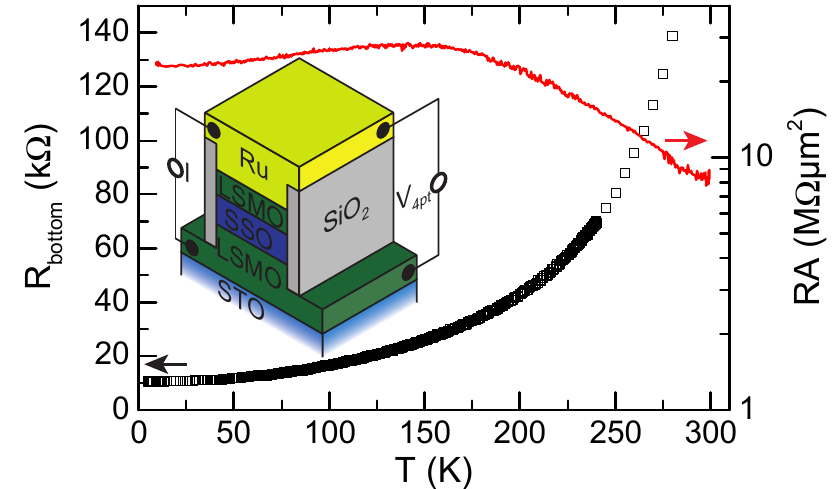}\\
  \caption[temperature dependence of resistance]{(Color online) Temperature dependence of the bottom contact resistance (black symbols) and of the MTJ resistance-area product (red line). The resistance of the bottom contact decreases with decreasing temperature. In contrast, the resistance-area product of the MTJ increases with decreasing $T$ (note the logarithmic scale), indicating that the resistance of the junction is dominated by the tunnel barrier and not the contact leads. The RA has been obtained under a bias voltage $V_\mathrm{4pt}=100\;\mathrm{mV}$.
  \label{figure:RT_LSMO_SSO}}
\end{figure}

In the following we present our electrical transport data, focusing on the results obtained for a LSMO(43)/SSO(2)/LSMO(25) trilayer and a MTJ contact with a $2.5\times2.5\;\mathrm{\mu m^2}$ area. Note however, that similar results have been obtained for MTJs up to an area of $16\;\mathrm{\mu m^2}$. A schematic drawing of the measurement setup is shown in the inset of Fig.~\ref{figure:RT_LSMO_SSO}. We first discuss the temperature dependence of the resistance-area product (RA) of the MTJ obtained for a bias voltage $V_\mathrm{4pt}=100\;\mathrm{mV}$ as depicted in Fig.~\ref{figure:RT_LSMO_SSO}. Clearly, the RA increases with decreasing temperature $T$, while the LSMO bottom contact resistance measured simultaneously decreases with decreasing $T$. From this finding we conclude that the resistance of the MTJ is dominated by the tunnel barrier itself. In addition, the resistance of the junction is at least one order of magnitude larger than the bottom contact such that any magnetoresistance contributions from the bottom LSMO contact can be neglected in our measurements. Interestingly, the RA of MTJs based on a SSO barrier is rather large, reaching values of up to $30\;\mathrm{M\Omega\mu m^2}$ compared to values of up to $500\;\mathrm{k\Omega\mu m^2}$ obtained for the STO based reference junctions at $V_\mathrm{4pt}=100\;\mathrm{mV}$. We attribute the increase to the larger bandgap of SSO ($4.1\;\mathrm{eV}$, Ref.~\onlinecite{Zhang2006174}) as compared to STO ($3.3\;\mathrm{eV}$, Ref.~\onlinecite{STO_bandgap_2001}), which directly leads to an increase in junction resistance. However, we would like to note that the high defect density in the SSO barrier might also play a crucial role for the RA, but would require further systematic studies.

\begin{figure}[b,t]
  \includegraphics[width=85mm]{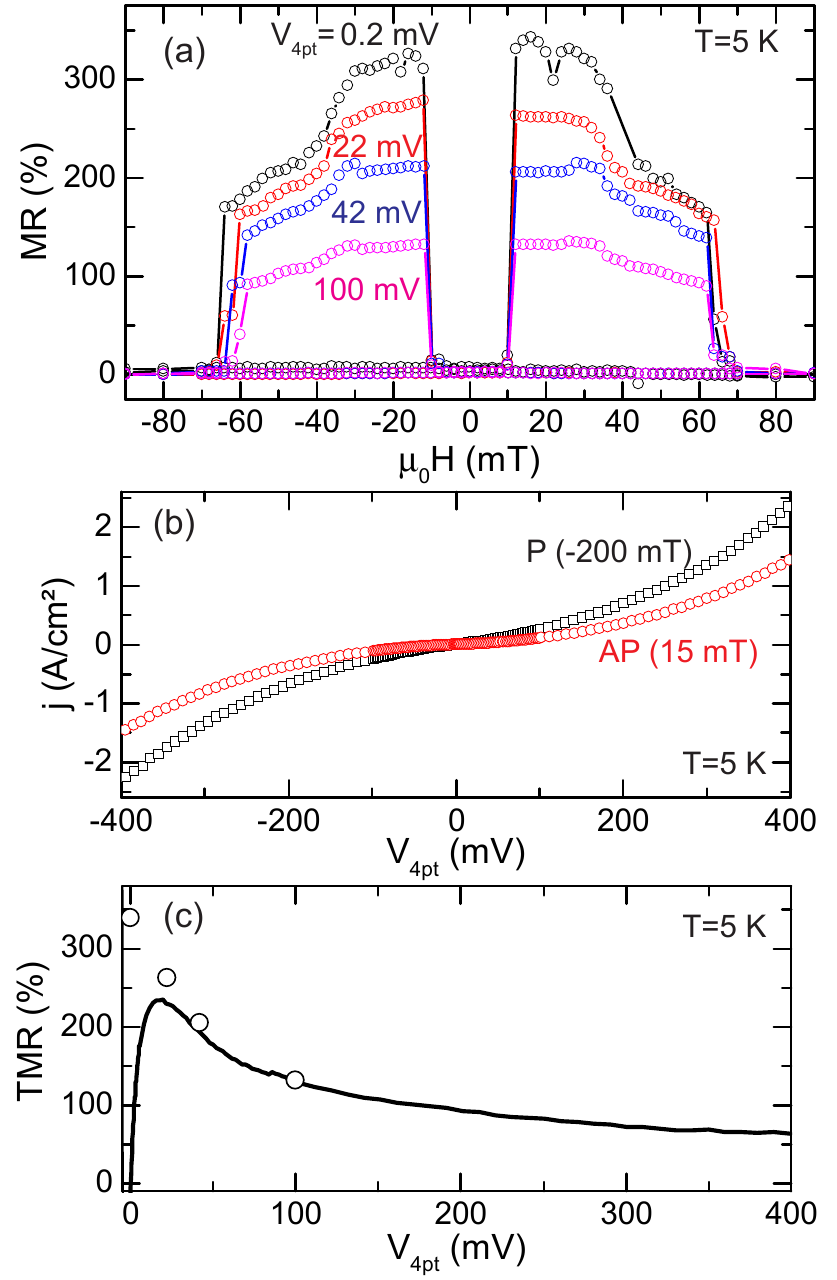}\\
  \caption[I-V characteristics at low T]{(Color online) (a) Bias dependence of the magnetoresistance at $T=5\;\mathrm{K}$ for a bias voltage of $0.2\;\mathrm{mV}$ (black), $22\;\mathrm{mV}$ (red), $42\;\mathrm{mV}$ (blue), and $100\;\mathrm{mV}$ (magenta) in a $2.5\times2.5\;\mathrm{\mu m^2}$ sized junction. The MR decreases with increasing bias voltage. Arrows in the figure indicate the sweep direction of the external magnetic field. (b) Current density vs applied bias voltage at $T=5\;\mathrm{K}$ for the very same sample as in (a) at two different external magnetic field values. For the electrode magnetization in the parallel state at $\mu_0 H=-200\mathrm{mT}$ (black symbols) a higher current flows as compared to the antiparallel arrangement at $\mu_0 H=15\mathrm{mT}$(red symbols). (c) Bias voltage dependence of the TMR extracted from the current-voltage characteristics of (b) (black line) and from the magnetic field sweeps at different $V_\mathrm{4pt}$ of (a).
  \label{figure:IV_LSMO_SSO}}
\end{figure}
As a next step we look into the magnetic field-dependent magnetoresistance (MR) of the $2.5\times2.5\;\mathrm{\mu m^2}$ SSO junction at $T=5\;\mathrm{K}$. For this we measured $I(H)$ at fixed $V_\mathrm{4pt}$ while sweeping the external applied magnetic field $H$ applied in the film plane along the bottom electrode strip from $\mu_0 H=-200\mathrm{mT}$ to $\mu_0 H=200\mathrm{mT}$ (upsweep) and back to $\mu_0 H=-200\mathrm{mT}$ (downsweep). From these measurements we calculated first the resistance $R(H)=V_\mathrm{4pt}/I(H)$ and then the magnetoresistance $MR(H)$ using
\begin{equation}
MR(H)=\frac{R(H)-R(-200\;\mathrm{mT})}{R(-200\;\mathrm{mT})}.
\label{equ:MR}
\end{equation}
The results of this procedure for different bias voltages $V_\mathrm{4pt}$ are compiled in Fig.~\ref{figure:IV_LSMO_SSO}(a). For all bias values used, we observe the typical pseudo spin-valve behaviour of a MTJ with two free magnetic layers: For large negative magnetic fields the magnetizations of the two LSMO electrodes are oriented parallel, which results in a low resistance state. When the external magnetic field is then increased to $\mu_0 H=10\;\mathrm{mT}$ the bottom LSMO electrode changes its orientation, resulting in an (nearly) antiparallel alignment of the two magnetizations. We observe a high resistance state, yielding a positive MR. We verified that the first switching is indeed coming from the bottom electrode by an independent anisotropic magnetoresistance measurement of the bottom electrode. We find a nice agreement between the switching fields extracted from both independent measurements (data not shown here). By further increasing $H$ the magnetization of the top LSMO layer also changes its orientation resulting again in a parallel orientation of the two magnetizations and a low resistance state at $\mu_0 H=670\;\mathrm{mT}$. A similar behaviour is observed when reversing the sweep direction only at negative fields. It is quite remarkable that bottom and top LSMO layer exhibit different switching fields. We attribute this to the difference in strain of the two layers induced by either the lattice mismatch between LSMO and SSO or the fabrication of the MTJs itself and the therefore resulting difference in lateral size for the top and bottom electrode. The difference in magnetic switching fields is further highlighted by the much steeper MR change occurring when the bottom electrode changes its orientation around $\mu_0 H=10\;\mathrm{mT}$ and the more gradual change in MR when the top electrode starts to change into the parallel configuration for $30\;\mathrm{mT} \leq \mu_0 H \leq 70\;\mathrm{mT}$. This may be explained by higher volume to surface ratio of the top LSMO electrode making it more susceptible to pinning defects, due to surface roughness, introduced by the MTJ fabrication process. In an independent series of experiments we measured the coercive fields of blanket LSMO films grown on (001)-oriented STO substrates with and without a $10\;\mathrm{nm}$ thick SSO layer between LSMO and substrate using a vibrating sample magnetometer (data not shown here). Within the experimental error we could not find any difference in the coercive fields for the different LSMO layers. These results suggest that the fabrication of the MTJs is an important factor for explaining the differences in switching fields for the top and bottom LSMO electrodes and further supports our assumption that pinning occurs due to the surface roughness introduced via the MTJ fabrication process.

We also studied the effect of the voltage bias on the observed $MR(H)$ for $V_\mathrm{4pt}=0.2,22,42, 100\;\mathrm{mV}$ as shown in Fig.~\ref{figure:IV_LSMO_SSO}(a) encoded in the graph as black, red, blue, and magenta symbols, respectively. The observed maximum MR increases with decreasing $V_\mathrm{4pt}$. Such a behaviour should be expected if spin-flip scattering across the tunnel barrier and magnon scattering is enhanced for higher bias values.~\cite{LeClair2005} In addition, the coercive field of the top electrode shifts to lower magnetic fields when increasing $V_\mathrm{4pt}$. We attribute this to a slight temperature increase of the MTJ due to the higher current density flowing through it at larger voltage bias. Another possible explanation may be spin transfer torque due to the not perfectly antiparallel orientation of bottom and top electrode. However, to quantify this effect, further experiments have to be conducted, which go beyond the scope of this work.

To get a deeper insight into the bias dependence we also recorded I-V curves at a fixed applied magnetic field for parallel ($-200\;\mathrm{mT}$, black symbols) and antiparallel ($15\;\mathrm{mT}$) alignment of the ferromagnetic electrodes by changing $V_\mathrm{4pt}$ and recording $I$ as shown in Fig.~\ref{figure:IV_LSMO_SSO}(b). Comparing the two different I-V curves, it becomes apparent, that for antiparallel alignment the MTJ is in its high resistance state (lower current density for same bias voltage), while in the parallel state it is in its low resistance state (higher current density for same bias voltage). This finding is in agreement with the standard Julliere two spin current model and the difference in scattering rates due to different density of states at the fermi level for majority and minority spin carriers~\cite{Julliere1975}. Moreover, we applied a Simmons fit~\cite{Simmons1963} to the parallel magnetization state, which yields $3.2\;\mathrm{eV}$ barrier height and $2.6\;\mathrm{nm}$ barrier thickness for SSO. The barrier thickness agrees reasonably well with the $2\;\mathrm{nm}$ determined from X-ray reflectometry measurements. The determined barrier height is lower than the bulk bandgap of SSO ($4.1\;\mathrm{eV}$~\cite{Zhang2006174}), which we attribute to the large concentration of defects in our SSO layer as observed in STEM images (see Fig.~\ref{figure:TEM_LSMO}(a)). Moreover, the barrier height and bandgap of the material are not one and the same quantity: The barrier height will also depend on the position of the Fermi energy of the contact electrodes with respect to the bandgap, such that the barrier height should always be smaller than the band gap of the barrier material.

Going a step further we used the obtained I-V curves to directly calculate the bias dependence of the TMR (maximum of MR in Fig.~\ref{figure:IV_LSMO_SSO}(a)) by determining the resistance for each bias voltage in the parallel ($R_\mathrm{para}(V_\mathrm{4pt})$ at $-200\;\mathrm{mT}$) and antiparallel ($R_\mathrm{anti}(V_\mathrm{4pt})$ at $15\;\mathrm{mT}$) alignment and using these values to calculate the TMR via
\begin{equation}
TMR(V_\mathrm{4pt})=\frac{R_\mathrm{anti}(V_\mathrm{4pt})-R_\mathrm{para}(V_\mathrm{4pt})}{R_\mathrm{para}(V_\mathrm{4pt})}.
\label{equ:TMR}
\end{equation}
The obtained TMR bias dependence is shown in Fig.~\ref{figure:IV_LSMO_SSO}(c) drawn as a black line. In addition, we included the MR obtained from the full $MR(H)$-loops [calculated also by Eq.~(\ref{equ:TMR})] for the 4 different $V_\mathrm{4pt}$ (Fig.~\ref{figure:IV_LSMO_SSO}(a)) as black circles into Fig.~\ref{figure:IV_LSMO_SSO}(c). We obtain a good agreement between the two differently determined TMR values for $V_\mathrm{4pt}\geq 42\;\mathrm{mV}$. For lower bias values both methods yield different results: While for the TMR determined from I-V curves we see a decrease of the TMR for low bias values, the TMR determined from $MR(H)$ loops monotonically increases with decreasing $V_\mathrm{4pt}$. We attribute these discrepancies to small temperature fluctuations during the measurement of the I-V curves, which are most prominent at small bias values, leading to the observed bias dependence. Thus, we assume that we in reality observe an increase of the TMR with decreasing $V_\mathrm{4pt}$. As already mentioned above, this behaviour is observed, when spin-flip and magnon scattering play a dominant role and influences on the TMR due to the band structure of the two ferromagnets can be neglected. A similar bias dependence has been observed in our STO based reference MTJs and also in other publications with STO as a tunnel barrier~\cite{lu_large_1996}. The largest TMR value of $350\;\%$ in our junction for $T=5\;\mathrm{K}$ is obtained at $V_\mathrm{4pt}= 0.2\;\mathrm{mV}$. Similar values have already been obtained in LSMO based MTJs with STO as a tunnel barrier, for example Garcia et al.~\cite{garcia_temperature_2004} reported a maximum TMR of $540\;\%$ at $4.2\;\mathrm{K}$, while Sun et al.~\cite{Sun1996} reported a maximum TMR of $\approx200\;\%$ at $4.2\;\mathrm{K}$. For LSMO junctions with STO barriers the TMR could be enhanced by up to $1900\;\%$ at low temperatures by carefully optimizing growth, layer structures and patterning process of the MTJs~\cite{Bowen2003,Werner2011}. This suggests that the TMR obtained in LSMO based MTJs with a SSO barrier could be further enhanced by further optimization. However, the TMR in these junctions is already comparable to values obtained with a STO barrier, despite the large density of defects present in the SSO barrier as determined by TEM analysis.

\begin{figure}[t,b]
  \includegraphics[width=85mm]{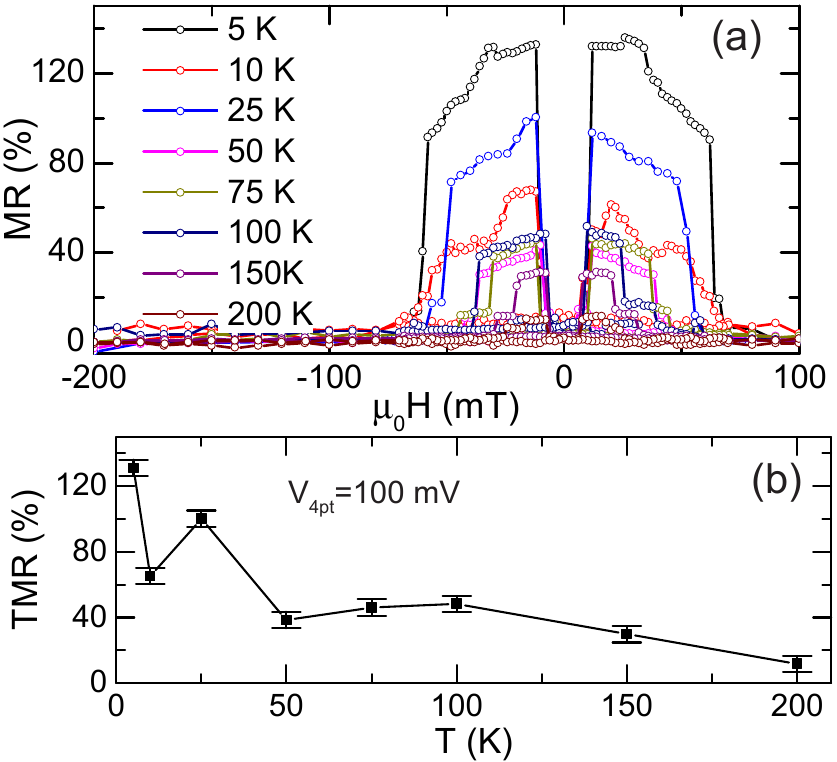}\\
  \caption[TMR as a function of T]{(Color online) Temperature dependence of the TMR at fixed bias voltage $V_\mathrm{4pt}= 100\;\mathrm{mV}$. (a) $MR(H)$ loops obtained at $T=5,10,25,50,75,100,150,200\;\mathrm{K}$, depicted with different coloured circles (see legend in the graph). (b) Temperature dependence of the TMR extracted from $MR(H)$. The TMR decreases with increasing $T$, for $T>200\;\mathrm{K}$ the TMR completely vanishes.
  \label{figure:TMRTemp_LSMO_SSO}}
\end{figure}

We investigated the temperature dependence of the TMR by measuring $I(H)$ loops at fixed $V_\mathrm{4pt}= 100\;\mathrm{mV}$ and $T$. From these measurements we determined the $R(H)$ and $MR(H)$ via Eq.\eqref{equ:MR}, the results of this procedure are depicted in Fig.~\ref{figure:TMRTemp_LSMO_SSO}(a). The MR decreases with increasing temperature, for temperatures larger than $200\;\mathrm{K}$ no MR was visible within the noise limit of our setup. The vanishing of the MR at high temperatures can be attributed to a possible dead magnetic LSMO layer at the interface between LSMO and SSO, similar as already discussed in literature for LSMO/STO interfaces.~\cite{Sun1999,Bibes2001} Moreover, the switching field of the top LSMO electrode decreases from $\mu_0H=60\;\mathrm{mT}$ at $T=5\;\mathrm{K}$ to $\mu_0H=24\;\mathrm{mT}$ at $T=200\;\mathrm{K}$. In contrast, the switching field of the bottom electrode changes only slightly from $\mu_0H=10\;\mathrm{mT}$ at $T=5\;\mathrm{K}$ to $\mu_0H=6\;\mathrm{mT}$ at $T=200\;\mathrm{K}$. We attribute this different evolution of the switching fields with temperature, as already discussed above, to the difference in strain and contributions of magnetic pinning defects for the top and bottom LSMO layer. The contributions of both effects will be reduced when increasing the temperature, leading to a reduction of the switching field of the top LSMO electrode.

From the $MR(H)$-loops at different $T$ we extracted the TMR as a function of temperature by taking the maximum value of each $MR(H)$ curve at each different $T$. The resulting TMR temperature dependence is depicted in Fig.~\ref{figure:TMRTemp_LSMO_SSO}(b). Overall, the TMR decreases with increasing $T$ and vanishes for $T>200\;\mathrm{K}$. However, the TMR extracted for $T=10\;\mathrm{K}$ clearly deviates from the general monotonic decreasing trend and is significantly lower than for $T=25\;\mathrm{K}$. We attribute this difference to the fact that the $T=10\;\mathrm{K}$ data has been obtained in a second cooling cycle, which might have changed the magnetic domain configuration at the interface.~\cite{Werner2011}

\section{Conclusions}
In summary, we have investigated for the first time the measurement of TMR in MTJs consisting of LSMO electrodes and SSO as the barrier. Our results suggest that SSO could be a promising barrier material for MTJs, with quite good insulating behaviour even when a high density of defects is present in the barrier. Our results show that the observed TMR in LSMO based MTJs with SSO as a barrier is comparable to the results obtained with STO as a barrier. However, further improvements in the SSO barrier properties, especially reducing the number of defects by lowering the lattice mismatch between the magnetic electrodes and the SSO barrier, are necessary to fully explore the potential of SSO as a barrier for MTJs. A very promising approach might be the usage of better lattice matched magnetic electrode materials, for example certain full Heusler compounds. For Co$_2$FeAl a very small lattice mismatch is obtained as $\sqrt{2}a_\mathrm{SSO}=0.5693\;\mathrm{nm}$ is extremely close to the bulk lattice constant of Co$_2$FeAl ($a_\mathrm{Co_2FeAl}=0.569\;\mathrm{nm}$ Ref.~\cite{Ebke2010}). We thus expect SSO to represent a very interesting material choice for further MTJ experiments.
\begin{acknowledgments}
We gratefully acknowledge financial support via NSF-ECCS Grant No.~1102263. Work at ORNL was supported by the U.S.~Department of Energy (DOE) Office of Science, Office of Basic Energy Sciences, Materials Science and Engineering Directorate.
\end{acknowledgments}
\bibliography{Biblio}
\end{document}